\documentclass[a4paper,floatfix,superscriptaddress,pra,twocolumn,showpacs,notitlepage,longbibliography,nofootinbib,aps,10pt]{revtex4-1}
\usepackage[T1]{fontenc}
\usepackage{dsfont}
\usepackage{tikz}
\usepackage{float}
\usepackage{mathrsfs}
\usepackage{amsmath}
\usepackage{amssymb}
\usepackage{graphicx}
\usepackage{cases}
\usepackage{esint}

\usepackage{mleftright} % For adequate spacing in formulas
\usepackage[unicode=true,
 bookmarks=false,
 breaklinks=false,pdfborder={0 0 1},backref=section,colorlinks=false]
 {hyperref}

\usepackage[english]{babel}

\usepackage{stmaryrd} %bracket

\makeatletter
\@ifundefined{textcolor}{}
{%
 \definecolor{BLACK}{gray}{0}
 \definecolor{WHITE}{gray}{1}
 \definecolor{RED}{rgb}{1,0,0}
 \definecolor{GREEN}{rgb}{0,1,0}
 \definecolor{BLUE}{rgb}{0,0,1}
 \definecolor{CYAN}{cmyk}{1,0,0,0}
 \definecolor{MAGENTA}{cmyk}{0,1,0,0}
 \definecolor{YELLOW}{cmyk}{0,0,1,0}
}

\usepackage{babel}

\DeclareMathOperator{\tr}{Tr}

\usepackage[braket, qm]{qcircuit}

\usepackage[normalem]{ulem}

\newcommand{\ignore}[1]{}
\newcommand{\sub}[2]{{\color{red}\sout{ #1}}{\color{blue}#2}}

\newcommand{\ad}[1]{{\color{blue} #1}}

\newcommand{\ex}[1]{{\color{red} \sout{#1}}}

\usepackage{comment}

\makeatother

\begin{document}
\title{Lie groups for quantum complexity and barren plateau theory}
\author{P. A. S. de Alcântara}
\email{pedro.antonio.alcantara@usp.br}
\affiliation{ICMC, University of S\~ao Paulo, PO Box 369,
13560-970, S\~ao Carlos, SP, Brazil}

\author{Gabriel Audi}
\email{gabriel.audi100@gmail.com}
\affiliation{Instituto de Física de São Carlos, Universidade de São Paulo, CP 369, 13560-970 São Carlos, SP, Brazil}

%\author{Reginaldo de Jesus Napolitano}
%\affiliation{S\~ao Carlos Institute of Physics, University of S\~ao Paulo, PO Box 369,
%13560-970, S\~ao Carlos, SP, Brazil}
\author{Leandro Morais}
\email{leandro.silva@ifsc.usp.br}
\affiliation{Instituto de Física de São Carlos, Universidade de São Paulo, CP 369, 13560-970 São Carlos, SP, Brazil}
\affiliation{International Institute of Physics, Federal University of Rio Grande do Norte, 59078-970 Natal, Brazil}

%\author{Pedro Rios} 
%\affiliation{ICMC, University of S\~ao Paulo, PO Box 369,
%13560-970, S\~ao Carlos, SP, Brazil}

\begin{abstract}
Advances in quantum computing over the last two decades have required sophisticated mathematical frameworks to deepen the understanding of quantum algorithms. In this review, we introduce
the theory of Lie groups and their algebras to analyze two fundamental problems in quantum computing as done in some recent works, e.g. \cite{nielsengeometriclowerbound,cerezo2021variational}. Firstly, we describe the geometric formulation of quantum computational complexity, given by the length of the shortest path on the $SU(2^n)$ manifold with respect to a right-invariant Finsler metric. Secondly, we deal with the barren plateau phenomenon in Variational Quantum Algorithms (VQAs), where we use the Dynamical Lie Algebra (DLA) to identify algebraic sources of untrainability.% This paper demonstrates that Lie theory provides a unifying framework to analyze quantum complexity and VQA trainability.
\end{abstract}

\maketitle

\section{Introduction}
Quantum Mechanics is the physical theory that describes nature at atomic and subatomic scales, where Classical Mechanics is no longer consistent with experimental results \cite{FeynmanVol3}. As a foundational theory, its principles have %not only led to the development of advanced frameworks like
led to the development of many areas such as
Quantum Field Theory \cite{peskin2018introduction}, the theoretical basis for the Standard Model of particle physics \cite{cottingham2007introduction}, %but have also spurred the emergence of new areas, most notably
and
Quantum Information \cite{nielsen_chuang}. Across these domains, its principles have been rigorously tested, consistently demonstrating remarkable success in predicting and describing physical phenomena.

The {growth} of quantum theory enabled the development of what is now recognized as the \textit{first generation of quantum technologies} \cite{georgescu2012quantum}. Spanning from the 1940s to the 1990s, this era was defined by innovations harnessing phenomena such as quantum tunneling and particle spin. Early developments included nuclear magnetic resonance (NMR) in the 1940s~\cite{PhysRev.70.460,PhysRev.69.37}, followed by tunneling Esaki diodes in the 1950s~\cite{PhysRev.109.603}, and Josephson junctions with superconducting quantum interference device (SQUIDs) in the 1960s~\cite{josephson1962possible}. This wave of innovation continued into the 1980s and 1990s with the rise of two crucial fields: spintronics~\cite{vzutic2004spintronics}, which led to magnetic sensors, hard disk drives, and magneto-resistive random access memory (MRAM); and quantum-dot technologies~\cite{reed1993quantum}, which have found applications ranging from medical imaging and photovoltaics to quantum computation.

Subsequently, beginning in the 1990s, a \textit{second generation of quantum technologies} emerged \cite{Shor,cirac1995quantum,bennett1992quantum}, leveraging more subtle quantum-mechanical phenomena such as superposition, entanglement, and more recently nonstabilizerness \cite{howard2014contextuality}. This new wave encompasses developments in quantum computation, quantum communication, and quantum sensing. These technologies utilize quantum resources to process information in fundamentally new ways, offering the potential for revolutionary applications such as %perfectly
secure communication and the ability to solve computationally intractable problems.

In this context, quantum computing (QC) occupies a paradoxically prominent position within the quantum technologies ecosystem, a prominence not grounded in its current level of technological maturity but rather in its %immense
long-term disruptive potential. Representing a paradigm shift that introduces a new class of computability, QC offers the promise of solving complex problems that remain intractable even for the most powerful classical supercomputers.

To truly demonstrate the computational superiority for certain problems of quantum computation over classical, we must resort to quantum complexity theory \cite{qu_comp_inf}. This field of study provides mathematical proofs of the advantages and limitations of a quantum {algorithm}. The quantum complexity of an algorithm is defined as the minimum number of computational steps required to perform some computation, where these computational steps are now given by quantum logical gates \cite{quantumcompasgeo}. Analogously to the classical case, the efficiency of a quantum {algorithm} is measured by examining how its computational complexity depends on the number of input qubits. In this context, {an algorithm} is said to be efficient if its complexity grows at most polynomially as a function of the number of input qubits \cite{quantumcompasgeo}.

This disruptive potential is exemplified by transformative applications, such as the accurate simulation of molecules to accelerate the discovery of new drugs \cite{google2020hartree,kumar2024recent}, and the optimization of investment portfolios and risk analysis in the financial sector \cite{rebentrost2024quantum}. Additionally, its notoriety is amplified by the fundamental threat it poses to cybersecurity, since algorithms like Shor’s could, in theory, efficiently break public-key encryption systems such as RSA, which underpin modern digital security \cite{Shor}.

Thus, research in quantum computing is not merely a subject of intellectual curiosity but a foundational pillar for the advancement of humanity in the coming decades. This field is built upon rigorous theoretical principles, among which group theory stands out as a fundamental tool. It provides a conceptual bridge that translates abstract notions of symmetry into concrete computational operations, namely the quantum gates that act on states within a Hilbert space.

As in any closed quantum system, in ideal quantum computation, %
%
%In an ideal quantum computation,
given an initial quantum state, the time evolution is determined by the Schrödinger equation, where the Hamiltonian operator governs %both 
the system’s dynamics. %and its interactions.
Equivalently, this evolution can be represented by an unitary operator which %, acting on the initial state, generates the evolved state
can be interpreted as a quantum logical gate. 
A quantum algorithm is constructed as some sequential application of quantum gates from a set of universal operations.
Thus, an ideal quantum computation on $n$ qubits corresponds to an unitary transformation belonging to the special unitary Lie group $SU(2^n)$ \cite{quantumcompasgeo}.

To this end, the structure of the paper is as follows. Section II will provide
the necessary mathematical framework, reviewing concepts of manifolds, Lie groups, and Lie algebras, with a focus on compact groups such as $SU(n)$. Section III will 
apply these tools in two contexts: first, in the study of quantum complexity from the perspective of Riemannian and Finsler geometry, where complexity is 
associated with geodesics on the group; and, second, barren plateaus in deep
quantum circuits through the Dynamical Lie Algebra. Finally, Section IV will present the Discussions.

\section{Mathematical Framework} \label{sec_math}

The mathematical framework of this review relies mainly on Riemannian-Finsler Geometry and Lie Groups, so we need some basic notions about smooth manifolds. Aside for some technical topological details (see e.g. \cite{Lee_smooth_mani}), a \emph{$m$-dimensional manifold} $M$ is characterized by having an open covering $\mathcal O$ satisfying the following properties:
\begin{enumerate}
    \item For each $\mathcal U\in \mathcal O$, we have a homeomorphism\footnote{A continuous map with continuous inverse.}
    \begin{equation*}
    x: \mathcal{U} \subset M \to \mathcal{V} \subset \mathbb{R}^m\,,
\end{equation*}
called a \emph{chart}, inducing local coordinates $(x^1,...,x^m)$ on $M$ from $\mathbb R^m$.

\item For charts $x_k:\mathcal U_k\to \mathcal V_k$, $k=1,2$, the \emph{transition map}
\begin{equation*}
    x_1\circ x_2^{-1}:x_2(\mathcal U_1\cap \mathcal U_2)\to x_1(\mathcal U_1\cap \mathcal U_2)
\end{equation*}
is smooth as a map between open sets of $\mathbb R^m$.
\end{enumerate}
Intuitively, $M$ is constructed by gluing together open sets of $\mathbb R^m$ in a smooth way. In this setting, any map $f:M\to N$ between manifolds, with $\dim M = m$ and $\dim N = n$, is locally expressible using coordinates from $\mathbb R^m$ for the domain and coordinates from $\mathbb R^n$ for the codomain; with this, it is possible to talk about derivatives in the sense of Calculus: such map is \emph{smooth} if its coordinate representations are smooth. In particular, a \emph{diffeomorphism} is a smooth map with a smooth inverse.

That said, let $M$ be a $m$-dimensional manifold. We denote by $C^\infty(M)$ the space of smooth functions from $M$ to $\mathbb R$. It is a commutative algebra\footnote{An \emph{algebra} is a vector space $A$ equipped with a bilinear map product $A \times A \to A$.}  since linear combinations and products of smooth functions are smooth. For a given $p\in M$, there are different, but equivalent ways of defining the \emph{tangent space} of $M$ at $p$, denoted $T_pM$, we choose the more useful to us: $T_pM$ is the vector space of derivatives of $C^\infty(M)$ at $p$, that is, it is the collection of linear functionals $v:C^\infty(M)\to \mathbb R$ satisfying the Leibniz rule 
\begin{equation*}
    v(fg) = f(p) \, v(g) + v(f) \, g(p)
\end{equation*}
for every pair of functions $f,g \in C^\infty(M)$. One can prove that $T_pM$ has dimension $m$ just as the manifold $M$, and it is equivalent to the space of velocities of smooth curves in $M$ at $p$, so it is reasonable to call it the tangent space (cf. \cite{Lee_smooth_mani}). If $x:\mathcal U\subset M\to \mathcal V\subset \mathbb R^m$ is a chart with coordinates $(x^1,...,x^m)$, the set of partial derivatives
\begin{equation}\label{part_deriv}
    \left\{\left(\dfrac{\partial}{\partial x^k}\right)_p \ , \ \ k=1, \cdots , m \right\}
\end{equation}
is a basis of $T_pM$ for every $p\in \mathcal U$.

A vector field $X$ on $M$ is a choice of tangent vectors $X|_p\in T_pM$ for each $p\in M$ with smooth components with respect to the basis \eqref{part_deriv} of partial derivatives in local coordinates of $M$; the space of vector fields on $M$ is denoted by $\mathfrak X(M)$, and it is an algebra with the commutator of vector fields
\begin{equation}\label{lie_br}
    [X,Y](f) = X(Y(f))-Y(X(f)),
\end{equation}
for every $X,Y\in \mathfrak X(M)$ and $f \in C^\infty(M)$. %{\color{ForestGreen} O que eh uma Lie algebra?... Bilinear, Jacobi identity and $[X , X] = 0$. Tbm deixo a sugestão para definir a notação explicitamente. R: Eu defino álgebra de Lie mais pra frente como a álgebra dos campos invariantes à esquerda num grupo de Lie, evitando detalhes desnecessários sobre identidade de Jacobi etc. Aqui está sendo definido o colchete de Lie (ou comutador, se preferir) de campos vetoriais em uma variedade (um pequeno adendum: toda álgebra de Lie de dimensão finita é álgebra de Lie de um grupo de Lie -- vide teorema de Ado e terceiro Teorema de Lie --, então não tem nenhuma perda significativa no tratamento que tá sendo dado aqui).}

With this, we have a bonafide notion of pushing forward a tangent vector: for $v \in T_pM$ and $f: M\to N$, the \emph{pushforward} of $v$ by $f$ is the tangent vector $f_{\ast,p}(v)\in T_{f(p)}N$ given by
\begin{equation}
    f_{\ast,p}(v)(g) = v(g\circ f)
\end{equation}
for every $g \in C^\infty(N)$. It is straightforward to check that $f_{\ast,p}$ is a linear map from $T_pM$ to $T_{f(p)}N$. Furthermore, if $M=N$ and $f$ is a diffeomorphism, then we have a linear map $f_\ast:\mathfrak X(M)\to \mathfrak X(M)$ given by $f_{\ast,p}$ at every $p\in M$.

The disjoint union
\begin{equation}
    TM = \bigsqcup_{p\in M}T_pM
\end{equation}
of all tangent spaces of $M$ is its \emph{tangent bundle}. This space inherits from $M$ a structure of a $2m$-dimensional manifold: a chart $x:\mathcal U\subset M\to \mathcal V\subset \mathbb R^n$ induces a chart $\xi_x:T\mathcal U\subset TM\to \mathcal V\times\mathbb R^m\subset \mathbb R^{2m}$ on $TM$ by
\begin{equation}
    \xi_x(p,v) = (x(p),x_{\ast,p}(v))\, .
    \label{ind_chart_tang_b}
\end{equation}
The pushforward $x_{\ast,p}(v)$ is the same as writing $v$ in the basis \eqref{part_deriv}, 
\begin{equation}
    v = \sum_{k=1}^m v^k\, \left(\dfrac{\partial}{\partial x^k}\right)_p\implies x_{\ast,p}(v) = (v^1,...,v^m)\, ,
    \label{x_ast(v)}
\end{equation}
so that a vector field $X\in \mathfrak X(M)$ is equivalent to a smooth map $X:M\to TM:p\mapsto X|_p$ such that $X|_p\in T_pM$. The \emph{zero section} of $TM$ is the vector field on $M$ that sends every point $p$ to the null vector $0\in T_pM$. Denoting by $\boldsymbol 0\subset TM$ the image of the zero section, the space $TM\setminus \boldsymbol 0$ is the \emph{slit tangent bundle} of $M$.

\begin{comment}
The differential structure described so far provides a way to work with distances on $M$. A \emph{Riemannian metric} $g$ on $M$ is a choice of inner product on $TpM$ for every $p \in M$ such that, for any $X,Y\in \mathfrak X(M)$, the function
\begin{equation}
    M\to \mathbb R:p\mapsto g(X|_p,Y|_p)
\end{equation}
is smooth\footnote{We refer to \cite{Lee_rieman} for more details on the rich structures of Riemannian manifolds, here we will focus on the natural concept of distance that arises in this context.}. This smoothness condition can be stated coordinate-wise by asking that the functions $g_{k,l}:\mathcal U\to \mathbb R$,
\begin{equation}\label{coord_riem_metric}
     g_{k,l}(p) = g\left(\left(\dfrac{\partial}{\partial x^k}\right)_p,\left(\dfrac{\partial}{\partial x^l}\right)_p\right)\, ,
\end{equation}
is smooth for every chart $x:\mathcal U\subset M\to x(\mathcal U)\subset\mathbb R^m$ and induced basis \eqref{part_deriv}. If $M$ is endowed with a Riemannian metric, it is a \emph{Riemannian manifold}.

Let $M$ be a connected Riemannian manifold with metric $g$. Note that, for any $v\in T_pM$, the quantity $\sqrt{g(v,v)}$ is naturally the norm of $v$. Then the \emph{length} of a piecewise regular curve $\gamma:[a,b]\to M$ is given by
\begin{equation}\label{len_curve}
    \ell(\gamma) = \int_a^b \sqrt{g(\gamma'(t),\gamma'(t))}dt\, ,
\end{equation}
and the \emph{distance} $d(p,q)$ between $p,q \in M$ is the infimum of lengths of piecewise regular curves joining $p$ and $q$.
\end{comment}

Now, we will use the differential structure described so far to define some notion of distance on $M$. In what follows, we make the additional assumption that $M$ is a connected, and we will denote points of $TM$ by coordinates $(x,v)\equiv (x^1,...,x^m,v^1,...,v^m)$ induced by coordinates on $M$ as in \eqref{ind_chart_tang_b}-\eqref{x_ast(v)}. A \emph{Finsler structure} or \emph{Finsler metric} on a manifold $M$ \cite{Finsler_Bao,rund_finsler} is a globally defined function
\begin{equation*}
    F : TM \rightarrow [0, \infty)
\end{equation*}
with the following properties:
\begin{enumerate}
    \item \emph{Regularity}: $F$ is smooth on the slit tangent bundle $TM \setminus \boldsymbol0$.
    
    \item \emph{Positive homogeneity}:
    \begin{equation*}
        F(x, \lambda v) = \lambda \, F(x, v) \ \ \ \forall\,\lambda>0\, .
    \end{equation*}
    \item \emph{Strong convexity}: The Hessian matrix with entries
    \begin{equation}
        g_{kl}(x,v) = \dfrac{1}{2} \dfrac{\partial^{2} F^2}{\partial v^{k} \partial v^{l}}(x,v)\, ,
        \label{Hessian matrix}
    \end{equation}
    is positive definite at every point of $TM \setminus \boldsymbol 0$.
\end{enumerate}

Straight out from definition, one can show that a Finsler structure satisfies the following properties:
\begin{enumerate}
    \item \emph{Positivity}:
    \begin{equation*}
        F(x, v) \geq 0 \quad \text{and} \quad F(x, v) = 0 \implies v = 0;
    \end{equation*}

    \item \emph{Triangle inequality}:
    $$F(x, v_1 + v_2) \leq F(x, v_1) + F(x, v_2);$$

    \item \emph{Energy relation}:
    \begin{equation*}
        F^2(x, v) = \sum_{k,l=1}^{m}g_{kl}(x,v) \, v^k \, v^l\, .
    \end{equation*}
\end{enumerate}

Given a piecewise regular curve $\gamma: [a,b] \subset \mathbb{R} \rightarrow M$, with almost everywhere well defined velocity $v(\lambda) = \gamma'(\lambda)$, the \emph{length} $\ell_{F}(\gamma)$ of this curve induced by a Finsler structure $F$ is defined by
\begin{equation*}
    \ell_{F}(\gamma) = \int_a^b \|{v(\lambda)}\|_{F,\gamma(\lambda)}  \, d\lambda\, ,
\end{equation*}
where we introduce the notation
\begin{equation*}
  \|{v(\lambda)}\|_{F,\gamma(\lambda)} = F\big(\gamma(\lambda), v(\lambda) \big) \, . 
\end{equation*}
The \emph{distance} $d(p, q)$ between two points $p, q \in M$ is defined as the infimum of the lengths $\ell_{F}(\gamma)$ over all piecewise regular curves $\gamma$ connecting the two points. To avoid unnecessary discussions on connection, in this work, we will use the term \emph{geodesic} to talk about curves that locally minimize distances -- for more details see e.g. \cite{Lee_rieman}. If the functions $g_{k,l}$ do not depend explicitly on the velocity, we get just Riemannian metric \cite{Lee_rieman}. A diffeomorphism $\phi: M \rightarrow M$ is said to be an \emph{isometry} if
\begin{equation*}
    \ell_{F}(\gamma) = \ell_{F}(\phi \circ \gamma)
\end{equation*}
for all curves $\gamma$. In particular, if
\begin{equation*}
F(x, v) = F\big(\phi(x), \, \phi_{\ast}(v)\big)\, ,
\end{equation*}
then $\phi$ is an isometry.

Finally, a \emph{Lie group} $G$ is an algebraic group endowed with a smooth manifold structure such that the maps of product and inversion,
\begin{equation*}
    G\times G\to G:(g,h)\mapsto gh \ \ \mbox{and} \ \ G\to G: g\mapsto g^{-1}\, ,
\end{equation*}
are smooth. For example, consider the space of all invertible matrices of order $n$ with complex entries $GL(n,\mathbb C)$: it is an open set of the space of all matrices of order $n$ with complex entries $M(n,\mathbb C)\simeq \mathbb C^{n^2}\equiv \mathbb R^{4n^2}$, so it is a manifold with a global chart, and the maps of multiplication and inversion are, respectively, a polynomial and a rational map on the global coordinates; therefore $GL(n,\mathbb C)$ is a Lie group.

%\sub{consider the set of all invertible linear maps of a complex vector space $V$ to itself, denoted by $GL(V)$, and by fixing a basis for $V$, one construct $GL(n,\mathbb C)$, defined as the set of all invertible complex matrices of order $n$, which can be shown to be a Lie group,}{

A \emph{left action} of a group $G$ on a manifold $M$ is a smooth map
\begin{equation*}
    G\times M\to M:(g,p)\mapsto \varphi_g(p):= g \cdot p
\end{equation*}
such that $\varphi_e(p) = p$ and $\varphi_g( \varphi_h(p) ) = \varphi_{gh}(p)$ for every $g,h \in G$ and every $p\in M$. By the differential structure on Lie groups, we can translate the action to an ``infinitesimal form'', as explained below.

A Lie group $G$ has a trivial action on itself given by the so-called \emph{left translation map}: for $g \in G$,
\begin{equation*}
    l_g:G\to G:h\mapsto l_g(h) =gh\, .
\end{equation*}
Similarly, there is the \emph{right translation}
\begin{equation*}
    r_g:G\to G:h\mapsto r_g(h) = hg \, .
\end{equation*}
A \emph{left-invariant vector field} on $G$ is a vector field $X\in \mathfrak X(G)$ such that $(l_g)_\ast(X) = X$. The set $\mathfrak g$ of all left-invariant vector fields is not only a vector space, but it is also a Lie algebra\footnote{An algebra $A$ is said to be a \emph{Lie algebra} with \emph{Lie bracket} $\llbracket \cdot , \cdot \rrbracket : A \times A \to A$, if this operation is bilinear, alternate ($\llbracket x,x\rrbracket = 0$ for $x \in A$) and satisfies the Jacobi identity ($\llbracket\, x \, , \, \llbracket\, y \, , \, z \rrbracket \rrbracket + \llbracket\, z \, , \, \llbracket\, x \, , \, y \rrbracket \rrbracket + \llbracket\, y \, , \, \llbracket\, z \, , \, x \rrbracket \rrbracket = 0$, for all $x, y, z \in A$).} associated to the Lie group, with Lie bracket given by the commutator of vector fields \eqref{lie_br}. There is a natural isomorphism $T_eG \simeq \mathfrak g$, sending each $v \in T_eG$ to the vector field given by $X|_g = (l_g)_{\ast,e}(v)$, so we import the Lie algebra structure from $\mathfrak g$ to $T_eG$ and identify both spaces, eventually referencing the tangent space $T_eG$ as the Lie algebra. For our previous example of Lie group, since $GL(n,\mathbb C)$ is open in the linear space $M(n,\mathbb C)$, its Lie algebra $\mathfrak{gl}(n,\mathbb C)$ is naturally identified with $M(n,\mathbb C)$, and it happens that the commutator of vector fields of $\mathfrak{gl}(n,\mathbb C)$ is equivalent to the commutator of matrices in $M(n,\mathbb C)$.

With this, we have the \emph{exponential map} $\exp:\mathfrak g \to G$ defined in the following way: for each $X\in \mathfrak g$, let $\gamma_X$ be the solution of the ODE
\begin{equation}\label{ODE_1-par_subg}
    \begin{cases}
        \gamma_X'(t) = X|_{\gamma_X(t)} \\
        \gamma_X(0) = e
    \end{cases}\, ,
\end{equation}
i.e., $\gamma_X$ has initial point in the identity of $G$ and its velocity is given by $X$; then
\begin{equation*}
    \exp(X) := \gamma_X(1)\, .
\end{equation*}
It is worth highlighting that, for any given $X$, the curve $\exp(tX)\equiv \gamma_X(t)$ is a $1$-dimensional abelian Lie subgroup of $G$. It can be shown that the exponential map is a local diffeomorphism from some open neighborhood of $0\in\mathfrak g$ to some open neighborhood of $e\in G$ \cite[Proposition 20.8]{Lee_smooth_mani}. For $GL(n, \, \mathbb C)$, the exponential map $M(n, \, \mathbb C)\to GL(n, \, \mathbb C)$ is precisely the exponential of matrices
\begin{equation*}
   \exp(t X) \equiv \sum_{k = 0}^\infty \,\dfrac{t^k}{k!} \,X^k\, ,
\end{equation*}
where $X^0 \equiv \mathds{1}$. Through straightforward calculation, one can verify that this definition satisfies \eqref{ODE_1-par_subg}, and that for every matrix Lie group, the exponential map coincides with the usual matrix exponential.

Now, we just need to note that any $X\in \mathfrak g$ induces a vector field $\widetilde X$ on $M$ by the initial velocities of the curves of the form $\exp(tX)\cdot p$:
\begin{equation}
    \widetilde X|_p = \dfrac{d}{dt}\bigg|_{t = 0}\exp(tX)\cdot p\, .
\end{equation}
Since the exponential map is a diffeomorphism from an open neighborhood of $0\in \mathfrak g$ to an open neighborhood of $e \in G$, the action of $G$ on $M$ can be studied locally in this ``infinitesimal'' set up of vector fields. Any basis of $\mathfrak g$ -- and, by extension, their induced vector fields on $M$ -- is a set of \emph{infinitesimal generators} of the symmetry given by $G$.

Here we are mainly interested in specific kinds of actions on vector spaces, the representations. A \emph{representation} of $G$ on $\mathbb C^n$ is a \emph{Lie group morphism} (i.e., smooth group homomorphism) $\rho: G\to GL(n,\mathbb C)$, so that any $g \in G$ can be \emph{represented} by a matrix $\rho(g)$. We say that a representation is \emph{unitary} if its image lies inside the Lie group of unitary square matrices $U(n)$, which means that each matrix $\rho(g)$ preserves the canonical inner product of $\mathbb C^n$.

Representations have a good notion of ``building blocks'': a representation is \emph{irreducible} (an \emph{irrep} for short) if the linear space where it is defined on has no non trivial invariant subspaces, and it is \emph{completely reducible} if it is a direct sum of irreducible representations. Then, an unitary representation is either an irrep or it is completely reducible.

The Lie groups that describe symmetries of quantum computing systems are compact. Compactness is a topological property concerning open covers, but we can state it by means of a useful fact (that follows from Peter-Weyl Theorem \cite{folland}): a \emph{compact Lie group} is a Lie subgroup of $U(n)$ for some $n$. In particular, a basic example of a compact Lie group is $U(n)$ itself, with Lie algebra
\begin{equation}\label{u(n)_alg}
    \mathfrak u(n) = \{X \in M(n,\mathbb C): X^\dagger = - X\}\, .
\end{equation} 
However, since Quantum Mechanics is invariant under global phases, we can restrict attention to the special unitary Lie group $SU(n)$, defined as the subgroup of unitary matrices of order $n$ whose determinant is equal to $1$. Its associated Lie algebra is the subspace of $\mathfrak{u}(n)$ with null trace matrices,
\begin{equation}\label{su(n)_alg}
    \mathfrak{su}(n) = \{X\in M(n,\mathbb C): X^\dagger=-X,\tr(X) = 0\}\, .
\end{equation}
Here, we emphasize that in Physics literature, one will often find the elements of the Lie algebras of (special) unitary groups with a factorized imaginary unit, $X = iH$, as e.g. in the context of symmetries in Quantum Field Theory, or also with a factorized minus imaginary unit, $X = - iH$, as it is traditionally done in Quantum Mechanics, so that, in both cases, $H$ is Hermitian, and often has the interpretation of a Hamiltonian of some quantum mechanical process. We discuss this subtlety in detail further in Section \ref{Coordenadas no SU}.

As the name suggests, a compact group is ``small'' enough to have an unique invariant measure $\mu$ with unit volume, the \emph{Haar measure} \cite{Lee_smooth_mani}. Using this measure, we have
\begin{equation}\label{haar_vol_unit}
     \int_G d\mu (g) = 1 \, ,
\end{equation}
\begin{equation}\label{haar_prod_inv}
    \int_G f(gh)d\mu(g) = \int_Gf(g)d\mu(g) = \int_Gf(hg)d\mu(g)\, ,
\end{equation}
\begin{equation}\label{haar_inv_inv}
    \int_Gf(g^{-1})d\mu(g) = \int_Gf(g)d\mu(g)\, ,
\end{equation}
whenever $f:G\to \mathbb C$ is an integrable function. The Haar integral for $U(1)$, for example, is just the normalized usual integral
\begin{equation*}
    \dfrac{1}{2\pi}\int_{-\pi}^\pi f(\theta)d\theta\, .
\end{equation*}

It turns out that, using the Haar integral, one can prove that any representation of a compact group can be taken as unitary \cite{folland}, so we only consider representations for compact groups as Lie groups morphisms to $U(n)$. A strong consequence of this fact is that every representation of any compact group can be constructed as direct sum of irreps. For a compact group $G$, the decomposition of tensor products of irreps into a direct sum of irreps is called \emph{Clebsch-Gordan series} of $G$. Such series are well known specially for $SU(2)$, as it is just the usual ``summation of spin''.

The Haar integral can also be used to prove that, for compact Lie groups, the exponential maps are surjective \cite{tao}, so every element of a compact group $G$ is of the form $\exp(X)$ for some $X \in \mathfrak g$. This is easy to see for $U(1)\simeq S^1$, where \eqref{u(n)_alg} becomes
\begin{equation*}
    \mathfrak u(1) = \{i\theta: \theta \in \mathbb R\}
\end{equation*}
and it is well known that each element of the group is given by $e^{i\theta}$ for some $\theta \in \mathbb R$.

The differential version of a representation of a Lie group is the map $\rho_\ast: \mathfrak g\to M(n,\mathbb C)$ given by
\begin{equation*}
    \rho_\ast(X) = \dfrac{d}{dt}\bigg|_{t = 0}\rho(\exp(tX))\, .
\end{equation*}
For an unitary representation of the group, the image of the induced Lie algebra representation lies inside $\mathfrak u(n)$. It is a matter of straightforward calculation to get that this is a linear map satisfying
\begin{equation}\label{lie_alg_morph}
    \rho_\ast([X,Y]) = \rho_\ast(X)\rho_\ast(Y)-\rho_\ast(Y)\rho_\ast(X)\, .
\end{equation}
A linear map $\mathfrak g \to M(n,\mathbb C)$ satisfying \eqref{lie_alg_morph} is called a \emph{Lie algebra representation}, and the notion of irrep is naturally applicable to Lie algebras just as for Lie groups. So each representation of a Lie group induces a representation of its Lie algebra. Under the special condition of the group being simply connected, this process of differentiating is a bijection and the representations of the group can be recovered from the representations of its Lie algebra by the exponential map \cite{Lee_smooth_mani}. The representations of Lie algebras are simpler to work out because they are just fancy linear maps, so we can use techniques from Linear Algebra. Another fortuitous fact is that $SU(n)$ is simply connected, which brings the question: are the representations of $\mathfrak{su}(n)$ easy to describe?

The key fact to better understand representations of Lie algebra is that the kernel $\ker(\phi)$ of a representation $\phi: \mathfrak g\to M(n,\mathbb C)$ is an \emph{ideal}, that is, it is a vector subspace such that $[X,Y]\in \ker(\phi)$ whenever $X$ or $Y$ are in $\ker(\phi)$. So it is convenient to restrict ourselves to special kinds of Lie algebras: a Lie algebra $\mathfrak g$ is \emph{simple} if it has only the trivial ideals $\{0\}$ and $\mathfrak g$ itself; it is \emph{semisimple} if it is the direct sum of simple Lie algebras. The main theorem of Representation Theory classifies all irreps of semisimple Lie algebras, see \cite{hump}. Again, $\mathfrak{su}(n)$ is simple, and its irreps are determined by $(n-1)$-tuples of non negative integers. For example, for $\mathfrak{su}(2)$, the irreps are identified by integers $k\ge 0$ and associated to spin systems, where $j = k/2$ is the spin number.

To recover $\mathfrak{u}(n)$ from $\mathfrak{su}(n)$, we just need to add the abelian ideal generated by $i\mathds 1$, which accounts for the idea that $SU(n)$ and $U(n)$ differ by global phases, so the irreps of $\mathfrak u(n)$ can be easily obtained from irreps of $\mathfrak{su}(n)$. In general, if a Lie algebra $\mathfrak g$ is a direct sum
\begin{equation}
    \mathfrak g = \mathfrak s\oplus\mathfrak a\, ,
\end{equation}
where $\mathfrak s$ is semisimple and $\mathfrak a$ is abelian, then it is called a \emph{reductive Lie algebra}. By opening up the semisimple factor $\mathfrak s$ as a direct sum of simple Lie algebras, one gets that a reductive Lie algebra $\mathfrak g$ is a direct sum of ideals,
\begin{equation}\label{decomp_red_lie_alg}
    \mathfrak g = \bigoplus_{k=0}^{d}\mathfrak g_k \, ,
\end{equation}
where $\mathfrak g_0 = \mathfrak a$ is the abelian ideal and each $\mathfrak g_k$ for $1\le k\le d$ is a simple Lie algebra. A standard fact is that, not only $\mathfrak u(n)$ is reductive, but every of its Lie subalgebra are reductive as well \cite{Hall_lie}.

Summing up, $SU(n)$ is a compact simply connected Lie group with simple Lie algebra, so its representations are, in some sense, fully understood, and the representations of $U(n)$ can be obtained from what we know for $SU(n)$.

\section{Applications}\label{geometria}

%\subsection{Quantum error-correction - Stabilizer codes} 

\subsection{Geometry of Quantum Complexity} 
\textbf{Quantum complexity theory} provides a mathematical framework for the precise study of the advantages and limitations of a quantum computation. Computational problems are classified according to their intrinsic hardness, and the central idea is to organize problems into various \emph{complexity classes}, depending on how difficult they are to solve using a quantum computer. The standard approach to defining and quantifying quantum complexity is known as \emph{gate complexity}. It begins by fixing a \emph{universal set of quantum gates}, and defines the complexity of a unitary operation or algorithm as the \emph{minimum number of elementary gates} from this set required to approximate the computation within a specified error tolerance. \cite{nielsen_chuang} This is a natural extension of the concept of computational complexity from classical to quantum algorithms.

Complexity classes are defined by analyzing how the complexity of an algorithm scales with the number of input qubits. For example, the class \textbf{BQP} (\emph{Bounded-Error Quantum Polynomial Time}) is defined as the set of decision problems that can be solved by a quantum algorithm in polynomial time, with error probability bounded above by a fixed threshold, often taken to be $1/3$. \cite{nielsen_chuang} When the number of required computational steps grows at most polynomially with the input size, the computation is often referred to as being \emph{efficient} \cite{quantumcompasgeo}.

Beyond the discrete gate-based definition, a continuous and geometric formulation of quantum complexity was introduced by Nielsen and collaborators as an alternative approach \cite{nielsengeometriclowerbound, geometryofquco, quantumcompasgeo}. In this framework, the special unitary group $SU(2^n) $ is equipped with a right-invariant Riemannian metric that assigns a cost to each direction in the Lie algebra $ \mathfrak{su}(2^n) $, reflecting the increasing physical difficulty associated with implementing multi-qubit operations \cite{quantumcompasgeo}. A quantum computation, described by a unitary transformation $U \in SU(2^n)$, has its complexity defined as the length of the shortest geodesic connecting the identity $\mathds 1$ to a final target unitary $U$. This distance yields both a lower bound on the exact gate complexity and an upper bound on the approximate gate complexity, and it establishes a polynomial equivalence between the geometric and gate-based formulations \cite{nielsengeometriclowerbound, geometryofquco, quantumcompasgeo}.

This geometric formulation enables the use of tools from variational calculus to study quantum complexity. The framework is still under active development and holds great potential. In particular, the search for more well-behaved cost metrics remains an active area of research. Brown and Susskind in \cite{PhysRevD.97.086015} have proposed alternative metric structures to address curvature-related issues. More recently, Auzzi and collaborators in \cite{Auzzi} have introduced even more progressive and gradual penalty functions. 

Notably, Brown in \cite{Brown2023} has recently derived a quantum complexity lower bound using the Bishop–Gromov theorem, a result from Riemannian geometry. By applying this bound to Nielsen's complexity geometry, he showed that for a broad class of cost functions, the typical complexity of a unitary grows exponentially with the number of qubits. This work accomplishes Nielsen’s original vision of using differential geometry as a rigorous tool to bound and understand quantum complexity.

%This geometric formulation enables the use of tools from variational calculus to study quantum complexity. The framework is still under active development and holds great potential. In particular, the search for more well-behaved cost metrics remains an active area of research. Brown and Susskind have proposed alternative metric structures to address issues related to negative curvature and to connect complexity with gravitational phenomena. More recently, Auzzi has introduced smoother penalty functions, aiming to provide a more gradual cost scaling and better analytical control over the geometric properties.

%Notably, Brown in \cite{Brown2023} has recently derived a quantum complexity lower bound using the Bishop–Gromov theorem, a result from Riemannian geometry. By applying this bound to Nielsen's complexity geometry, he showed that for a broad class of cost functions, the typical complexity of a unitary grows exponentially with the number of qubits. This work realizes Nielsen’s original vision of using differential geometry as a rigorous tool to bound and understand quantum complexity.

To develop these ideas, we begin by introducing two coordinate systems for the Lie group $SU(2^n)$, which induce two distinct representations of tangent vectors: the Hamiltonian representation and the Pauli representation. We then present the first geometric formulation of quantum complexity and solve the geodesic equation for a class of problems known as the Pauli geodesics~\cite{nielsengeometriclowerbound}.

% it is possible to define the logarithm of any element of $SU(2^n)$, but not in a unique way, leading to the notion of \emph{branch} of the logarithm.

%$\exp:\mathfrak{su}(2^n)\to SU(2^n)$ is a local diffeomorphism around $0\in \mathfrak{su}(2^n)$.

\subsubsection{Coordinates in the $SU(2^n)$}\label{Coordenadas no SU}
As stated in section \ref{sec_math}, the exponential map is surjective for compact Lie groups such as $SU(2^n)$; therefore, any element $U\in SU(2^n)$ can be expressed as $U = \exp(-iH)$ for some $iH\in \mathfrak{su}(2^n)$. However, it is not injective, so it is not a global diffeomorphism. Yet, by restricting its domain and codomain to maximal open sets $ \mathcal W\subset \mathfrak{su}(2^n)$ and $\mathcal U\subset SU(2^n)$ such that $ 0\in \mathcal W$, $\mathds 1\in \mathcal U$, and $\exp|_{\mathcal W}:\mathcal W\to \mathcal U$ is a diffeomorphism, with inverse $\ln:\mathcal U\to \mathcal W$,
%\begin{equation}
%\begin{aligned}
 %    \ex{\ln:\mathcal U} &\ex{\to \mathcal W\,} \\
  %   \ex{U} &\ex{\mapsto \ln(U) := \sum_{k=1}^{\infty} \, \dfrac{(-1)^{k-1}}{k}  \,\big[ U - \mathds 1\big]^{k}}
%\end{aligned}
%\end{equation}
\begin{equation}
    \ln(U) = \sum_{k=1}^{\infty} \, \dfrac{(-1)^{k-1}}{k}  \,\big( U - \mathds 1\big)^{k}\, ,
\end{equation} commonly referred to as the \textit{logarithm} in the context of Lie group theory. A possible precise characterization of these suitable open sets can be achieved by explicitly choosing the standard branch of the logarithm, and with this choice, we avoid unitaries with an eigenvalue $-1$ \cite{nielsengeometriclowerbound}. 

% will be addressed later, especially in the definition of the Pauli coordinates, where they are not taken to be infinitesimal neighborhoods.

%As with the complex logarithm, the matrix logarithm is a local diffeomorphism, but globally it is a many-to-one map. We follow the approach of Ref.~\cite{nielsengeometriclowerbound} and define the so-called $U$-adapted coordinates and the Pauli coordinates for the Lie group $SU(2^n)$. 
%The precise characterization of these suitable open sets will be addressed later, when we construct the coordinate system where such conditions become relevant.We closely follow reference \cite{nielsengeometriclowerbound} and define the so-called $U$-adapted coordinates and the Pauli coordinates for the Lie group $SU(2^n)$. % We closely follow reference  \cite{nielsengeometriclowerbound} and define the so-called $U$-adapted coordinates and the Pauli coordinates for the Lie group $SU(2^n)$. We 

We begin establishing a basis for this Lie algebra by defining the generalized Pauli matrices \cite{Stab_genera_Rigetti}. Assigning numerical indices to the standard $2 \times 2$ Pauli matrices and identity matrix, so that $\sigma_{0} \equiv \mathds 1$, $\sigma_{1} \equiv \sigma_{x}$, $\sigma_{2} \equiv \sigma_{y}$ and $\sigma_{3} \equiv \sigma_{z}$, the generalized Pauli matrices $\sigma_{k}$ are defined as
\begin{equation}\label{sigma_k}
    \sigma_{k} := \sigma_{m_{1}}  \otimes \sigma_{m_{2}}  \otimes  \dots \otimes  \sigma_{m_{n}} \, , 
\end{equation}
where $k$ is determined from $m_1,...,m_n\in \{0,1,2,3\}$ via
\begin{equation*}
    k = 4^0 \, m_1+4^1\, m_2+...+4^{n-1} \, m_n\, .
\end{equation*}

Now we use the fact that the set of generalized Pauli matrices multiplied by the imaginary unit $i$ 
\begin{equation*}
    \left\{i \, \sigma_{k}, \, \textrm{with}  \, k= 1, \, 2,   \dots , \, 4^{n} -1\right\},
\end{equation*} forms a basis for the Lie algebra $\mathfrak{su}(2^{n})$ \cite{nielsengeometriclowerbound, Stab_genera_Rigetti}. We highlight that the identity element $\sigma_{0} = \mathds 1$ is not included in the basis, since the operators in this algebra are traceless. Moreover, we recall that the imaginary unit $i$ is required because the generalized Pauli matrices $\sigma_k$ are Hermitian, while the elements of the Lie algebra $\mathfrak{su}(2^n)$ must be anti-Hermitian by definition. Therefore, any element $iH \in \mathfrak{su}\big(2^{n}\big)$ can be written as $iH = i \, r \cdot \sigma$, where, for compactness, we introduce the notation
\begin{equation*}
    r \cdot \sigma := \sum_{k=1}^{4^{n} -1} \, r^{k} \,\sigma_{k}
\end{equation*} for $r = (r^1,...,r^{4^n-1})\in \mathbb R^{4^n-1}$. %m\ad{With this, we can identify $\mathcal W$ with an open neighborhood of the origin $\mathcal V\subset \mathbb R^{4^n-1}$ by}
%\begin{equation*}
 %   \ad{\mathcal V \ni q \longleftrightarrow -i q\cdot \sigma \in \mathcal W\, .}
%\end{equation*}

%We recall from equations \eqref{ind_chart_tang_b} that, to fully determine a tangent vector $v \in TM$, we can specify the point $p \in M$ such that $v \in T_{p}M$, and also the components of this vector in a certain basis for the tangent space $T_{p}M$. Thus, in our context of the special unitary Lie group, given a smooth curve
%Recall that, to fully determine a tangent vector $v \in TM$, we need to specify the point $p \in M$ such that $v \in T_{p}M$, and also the components of this vector in a certain basis for the tangent space $T_{p}M$ (see e.g. \eqref{ind_chart_tang_b}-\eqref{x_ast(v)}). %Thus, in our context of the special unitary Lie group, given a smooth curve
%
%such that $U(0) = \mathds 1$, at each point $U(\lambda)$, we aim to describe and parameterize the corresponding velocity vector}
%\begin{equation}
%   \ex{v(\lambda):= \dfrac{dU(\lambda)}{d\lambda} \in T_{U(\lambda)}SU(2^n)\,
%\end{equation}
%associated to the curve. Figure \ref{tan.vec} illustrates the idea. We will present the $U$-adapted coordinates and the Pauli coordinates \cite{Brandt2009, nielsengeometriclowerbound} and introduce an alternative and more compact notation for the basis induced by these specific charts.

%\subsubsection{Hamiltonian representation}\label{sec_hamiltonian}

We start introducing the $U$-adapted coordinates and the Hamiltonian representation \cite{Brandt2009, nielsengeometriclowerbound}. Fixing an origin $U \in SU(2^{n})$, %\sub{we define:}
{we take}
\begin{equation*}
    {\mathcal{U}_{ \hspace{0.05cm}U} = \{W \, U: W \in \mathcal U\}}
\end{equation*}
{and define}
\begin{equation*}
\begin{aligned}
     \phi_{U}: \mathcal{V} \subset \mathbb{R}^{4^{n}-1} \ & \to \hspace{1 em}\mathcal{U}_{\,U} \subset SU(2^{n}) \\
     r \hspace{2 em} & \mapsto \hspace{1 em}  V= \exp(-i \, r \cdot \sigma ) \, U.
\end{aligned}    
\end{equation*}
%\ex{where $\mathcal V$ is identified with $\mathcal W$ via}
%\begin{equation*}
%    \ex{\mathcal V \ni r \longleftrightarrow -i r\cdot \sigma \in \mathcal W}
%\end{equation*}
%\ex{and}
%\begin{equation*}
%    \ex{\mathcal{U}_{ \hspace{0.05cm}U} = \mathcal{ U} \,U = \{\widetilde U \, U: \widetilde U \in \mathcal U\}\, .}
%\end{equation*}
The chart $x_{U}$ is given by $x_{U} = \phi^{-1}_{U}$, so that%\ex{:}
\begin{equation}
    \begin{aligned}
        x_{U} : \hspace{2.5 em} \mathcal{U}_{\,U} \subset SU(2^{N}) \hspace{1 em} & \to \   \mathcal V \subset \mathbb{R}^{4^{n}-1}\\
     V = \exp(-i  \,  r \cdot \sigma )\, U \ &\mapsto \ x_{U}(V) = r,
    \end{aligned}
\end{equation}
and the coordinates are given explicitly by
\begin{equation*}
  x_{U}^{k}(V) = \dfrac{ i \, \textrm{Tr}\Big\{\ln\big(V \, U^{\dagger} \big) \,  \sigma_{k}  \Big\}}{2^{n}},  
  \label{cartaUadp}
\end{equation*} where $\ln$ is the aforementioned logarithm. %Note that, to define these charts, it is sufficient to take a neighborhood where the exponential map is a diffeomorphism.
For each point $U\in SU(2^n)$, we will use the following notation for the $k$-th element of the basis induced by the chart $x_{U}$ at that point $U$: 
\begin{equation*}
  e^{\prime}_{k,U} \equiv \left(\dfrac{ \partial  }{\partial x_{U}^{k}}\right)_{ U}.  
  \label{vetorlinha}
\end{equation*}
Summarizing, the $U$-adapted coordinates are a type of local coordinate system where the origin is defined at point $U$ \cite{nielsengeometriclowerbound}. The idea is that, when we move from point to point, we adapt the coordinates to have the origin at the new point.% %\ex{We will denote by $S'$ the frame in which we take $U(\lambda)$-adapted coordinates for each point $U(\lambda)$ in the curve $U$. The frame $S^{\prime}$ has its origin at point $U(\lambda)$ for every point on the curve, and we will refer to it as a moving frame.} %\ex{In the tangent spaces of each point $U(\lambda)$, we denote the basis induced by this type of coordinate system as the basis of ``prime vectors'':}
%$$\ex{\Big\{\Vec{e}^{\hspace{0.1cm}\prime}_{k,U(\lambda)}, \hspace{0.1cm} \textrm{with} \ k=1, 2, \dots , 4^{N} -1 \Big\}.}$$

\begin{figure}[H]
	\begin{center}
	\includegraphics[scale=0.45]{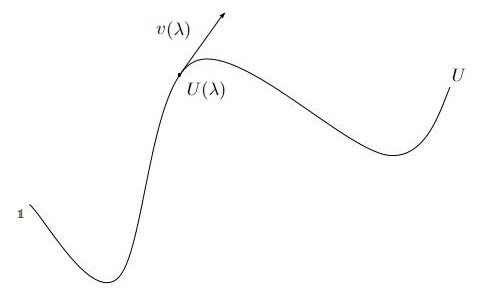} \\
    \caption{\label{tan.vec}The velocity $v(\lambda)$ of the curve $U$ at the point $U(\lambda)$.}
	\end{center}	
\end{figure}

Recall that, to fully determine a tangent vector $v \in TM$, we need to specify the point $p \in M$ such that $v \in T_{p}M$, and also the components of this vector in a certain basis for the tangent space $T_{p}M$ (see e.g. \eqref{ind_chart_tang_b}-\eqref{x_ast(v)}). That said, let
\begin{equation}\label{curve_U}
\begin{aligned}
    U: [0, \,  \infty) &\to SU(2^n)\\
    \lambda \ \ & \mapsto U( \lambda ),
\end{aligned}
\end{equation}
be a smooth curve such that $U(0) = \mathds 1$. At each point $U(\lambda)$, we aim to describe and parameterize the corresponding velocity vector
\begin{equation}
   v(\lambda):= \dfrac{dU(\lambda)}{d\lambda} \in T_{U(\lambda)}SU(2^n)
\end{equation} 
associated to the curve. Figure \ref{tan.vec} illustrates the idea.

%We will denote by $S'$ the frame in which we take $U(\lambda)$-adapted coordinates for every point in the curve. The frame $S'$ has its origin at point $U(\lambda)$ for every point on the curve, and we will refer to it as a \emph{moving frame}.}

We will denote by $S'$ the \emph{moving frame} in which we take $U(\lambda)$-adapted coordinates for every point of the curve. In this frame, we have the so-called Hamiltonian representation of the tangent vectors of this curve. To establish this representation, we note that, in first order in $\varepsilon$, we have
\begin{equation*}
\begin{aligned}
   U(\lambda + \varepsilon) &\approx  U(\lambda) + \varepsilon \frac{dU(\lambda)}{d\lambda}  \\[2ex]
   &= \left[\mathds 1 +  \varepsilon \frac{dU(\lambda ) }{d\lambda}  U^{\dagger}(\lambda)   \right]   U(\lambda) \\[2ex]
   &\approx \exp\left[-i \varepsilon \left(i \frac{dU(\lambda ) }{d\lambda} U^{\dagger}(\lambda)  \right)\right] U(\lambda).
\end{aligned}
\end{equation*}
Therefore, we can describe the curve by
\begin{equation*}
     U(\lambda + \varepsilon) \approx \exp\left[-i \varepsilon H(\lambda) \right] U(\lambda),
\end{equation*}
where
\begin{equation}\label{schr_eq}
     H(\lambda) = i \frac{dU(\lambda ) }{d\lambda} U^{\dagger}(\lambda)
\end{equation} 
is a Hermitian\footnote{Since $U(\lambda) \, U^{\dagger}(\lambda) = \mathds{1}$, it follows that 
\begin{equation*}
    \dfrac{d U(\lambda)}{d \lambda} \, U^{\dagger}(\lambda) = - U(\lambda) \, \dfrac{d U^{\dagger}(\lambda)}{d \lambda}\, ,
\end{equation*}
and $H^{\dagger}(\lambda) = H(\lambda)$ . } and traceless\footnote{From Jacobi’s formula,% given by 
\begin{equation*}
\frac{d}{d\lambda} \det \left[U(\lambda)\right] = \det\left[U(\lambda)\right]\tr\left[U^{-1}(\lambda) \, \dfrac{dU(\lambda)}{d\lambda} \right],
\end{equation*}  it follows that $\det\left[ U(\lambda)\right] = 1$ implies $\tr{\left[H(\lambda)\right]} = 0$.} operator satisfying the \emph{Schrödinger equation} in the Heisenberg picture.
%
%Now it is direct to see that this is precisely the \emph{Hamiltonian} operator of QM. 
%
For
\begin{equation*}
   h(\lambda) = \big(h^{1}(\lambda), \dots ,\, h^{4^{n}-1}(\lambda) \big) \in \mathbb{R}^{4^{n} -1}
\end{equation*}
given by
\begin{equation}\label{H-h_rel}
    H(\lambda) = h(\lambda) \cdot \sigma \, ,
\end{equation}
one can check from the definitions that the velocity of $U(\lambda)$ is given by
\begin{equation}
    v(\lambda) = \sum_{k=1}^{4^{n} -1} \, h^{k}(\lambda) \,  e^{\prime}_{k,U(\lambda)}.
\end{equation}

\begin{comment}
    
One can check from the definitions that the velocity of $U(\lambda)$ is given by
\begin{equation}
    v(\lambda) = \sum_{k=1}^{4^{n} -1} \, h^{k}(\lambda) \,  e^{\prime}_{k,U(\lambda)},
\end{equation} where
\begin{equation*}
   h(\lambda) = \big(h^{1}(\lambda), \dots ,\, h^{4^{n}-1}(\lambda) \big) \in \mathbb{R}^{4^{n} -1}
\end{equation*} are the components of the Hamiltonian in the generalized Pauli matrices basis, that is, $h(\lambda)$ is defined by
\begin{equation}\label{H-h_rel}
    H(\lambda) = h(\lambda) \cdot \sigma \, .
\end{equation}

\end{comment}

This is the Hamiltonian representation of this velocity vector. The curve's position is given at the origin of the local coordinate system, and the components of the tangent vector are given by the components of the Hamiltonian in the basis of generalized Pauli matrices.

%\ex{One can check from the definitions that the velocity tangent vector at each point $U(\lambda)$ of this curve, in the $S'$ frame, is given by}
%\begin{equation*}
  %  \ex{v(\lambda) = \sum_{k=1}^{4^{n} -1} \, h^{k}(\lambda) \,  e^{\prime}_{k,U(\lambda)},} 
%\end{equation*}
%\ex{where $h^{k}(\lambda) = \big(h^{1}(\lambda), \, h^{2}(\lambda), \dots , \, h^{4^{n}-1}(\lambda) \big) \in \mathbb{R}^{4^{n} -1} $ are the components of the Hamiltonian $H(\lambda)$ in the basis for of generalized Pauli matrices:}
%\begin{equation*}
   % \ex{H(\lambda) = h(\lambda) \cdot \sigma.} 
%\end{equation*}}
%\subsubsection{Pauli representation}\label{sec_Pauli}
Now, the Pauli coordinates are defined as $\mathds 1$-adapted coordinates. Denoting this chart by $x_{\mathds 1} \equiv x$, we have
\begin{equation}
    \begin{aligned}
        x\,:\hspace{2 em}   \mathcal{U} \subset  \textrm{SU}(2^{N})\hspace{1 em} &\to  \  \mathcal{V} \subset \mathbb{R}^{4^{N}-1}\\
     V= \exp(-i  \,  q \cdot \sigma ) \ &\mapsto \ \  x(V)=q
    \end{aligned}\, ,
\end{equation}
with%\ex{:}
\begin{equation*}
x^{k}(V) = \dfrac{ i \, \textrm{Tr}\Big\{\ln\big(V \big)  \, \sigma_{k}  \Big\}}{2^{n}}.
\label{cartaPauli}
\end{equation*}%\textcolor{red}{Aqui vem mais uma frase}
If a point $U$ lies in the domain of the chart $x$, we can induce a basis for the tangent space at the point. Thus, if $U \in \mathcal{U}$, we will use the following notation for the $k$-th element of the basis induced by the Pauli coordinates $x$ at that point $U$:
$$e_{k,U} \equiv \left(\dfrac{ \partial  }{\partial x^{k}}\right)_{ U}.$$ 

 %In the tangent spaces, for each point $U(\lambda)$, we have the basis of ``non-prime vectors'':
%$$ \Big\{  e_{k,U(\lambda)}, \, \textrm{with} \, k= 1,\, 2, \dots ,\, 4^{n} -1  \Big\}. $$

Pauli coordinates are a type of coordinate system that has its origin at the identity element of the group. Therefore, if every point $U(\lambda)$ of the curve \eqref{curve_U} lies in the domain of this coordinate system, we can establish a frame, denoted by $S$, in which we take Pauli coordinates for every point $U(\lambda)$ of the curve. This frame has its origin fixed at the identity $\mathds 1 \in SU(2^n)$ of the group; thus, we refer to it as a fixed frame. 

Now, we take
\begin{equation}
    q(\lambda) = (q^1(\lambda),...,q^{4^n-1}(\lambda))\in \mathbb R^{4^n-1}
\end{equation}
such that
\begin{equation}
    U(\lambda) = \exp(-i q(\lambda) \cdot \sigma)\, .
\end{equation}%
In first order in $\varepsilon$, we have:
\begin{align*}
   U(\lambda + \varepsilon) &=    \exp[-i \, q(\lambda +\varepsilon ) \cdot \sigma]  \\[1ex]
   &\approx \exp\big\{-i \, [ q(\lambda) + \varepsilon \, \dot{q}(\lambda)] \cdot \sigma \big\} .
\end{align*} In the frame $S$, one can check that the velocity at each point of this curve is given by:
\begin{equation*}
     v(\lambda) = \sum_{k=1}^{4^{n} -1} \, \dot{q}^{k}(\lambda) \, e_{k,U(\lambda)}.
\end{equation*}
This is the Pauli representation of this velocity vector. The curve's position is described by the coordinates $q(\lambda)$, while the velocity is described by $\dot{q}(\lambda)$. %and those are the parameters we will use in the Euler-Lagrange equations. % As the frame $S$ is fixed, $q(\lambda) $ and $\dot{q}(\lambda)$ are the parameters we will want to use in the Euler-Lagrange equations. We will then discuss the changes of basis between these coordinate systems. In figure we illustrate the idea of both reference frame $S$ and $S^{'}$.

%\subsubsection{Change of frame}\label{sec_change}
%

\begin{figure}[H]
	\begin{center}
	\includegraphics[scale=0.5]{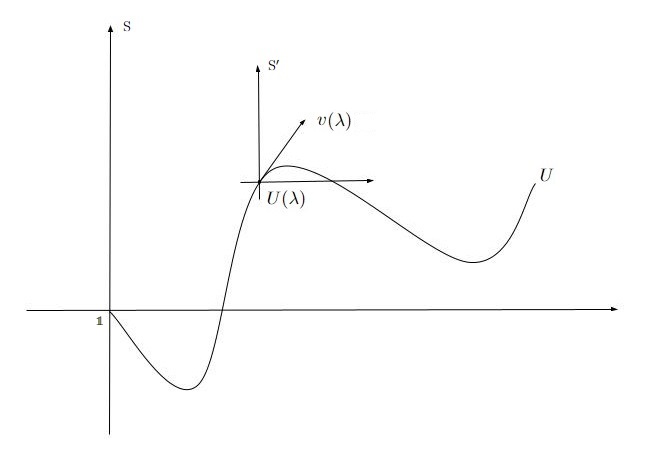} \\
    \caption{\label{fig_ref} The reference frame $S$ and $S^{'}$.}
	\end{center}	
\end{figure}

The change of frame from the Hamiltonian representation to the Pauli representation can generally be achieved by the Dyson series \cite{Reina_Quiroga, Mandl_Shaw}. Given some Hamiltonian $H(\lambda)$, the corresponding unitary operator, solution of the differential equation \eqref{schr_eq} with initial condition $U(0) = \mathds 1$, is formally and explicitly given by
\begin{equation*}
    U(\lambda) = \mathds 1 + \sum_{k=1}^{\infty} (-i)^{k} \int_{0}^{\lambda} d\lambda_{1} H(\lambda_{1}) 
  ...\int_{0}^{\lambda_{k-1}} d\lambda_{k} H(\lambda_{k})\, .
\end{equation*}
To write $U(\lambda)$, at least formally, as some exponential, one introduces the time-ordering operator $\mathcal{T}$ as
\begin{equation*}
    \mathcal{T}\left[ H(\lambda_1)\cdots H(\lambda_k) \right]
    = H(\lambda_{\alpha_1})\cdots H(\lambda_{\alpha_k}),
\end{equation*}
with $\lambda_{\alpha_1} > \lambda_{\alpha_2} > ... > \lambda_{\alpha_k}$. That is, the operator $\mathcal{T}$ reorders the product so that the parameters decrease from left to right. By permutation of the integration variables and the integration domains, we have
\begin{equation*}
     U(\lambda) = \mathds 1 + \sum_{k=1}^{\infty} \dfrac{(-i)^{k}}{k!} \int_{0}^{\lambda} d\lambda_{1}
    ... \int_{0}^{\lambda_{}} d\lambda_{k}   \mathcal{T}\left[ H(\lambda_1) ... H(\lambda_k) \right]
\end{equation*}
which motivates a shorthand notation for this expression, by defining the time-ordered matrix exponential as exactly that, to simply write
\begin{equation*}
    U(\lambda) = \mathcal{T}\exp\left(-i \int_{0}^{\lambda} \,  d\lambda^{\prime} \, H(\lambda^{\prime}) \right).
\end{equation*}

On the other hand, the change of frame from the Pauli representation $S$ to the Hamiltonian representation $S'$ is done through a version of the Baker–Campbell–Hausdorff formula \cite{Hall_lie}. Let $\big(q(\lambda), \dot{q}(\lambda)\big)$ be the Pauli coordinates of the velocity of the curve $U(\lambda)$. Defining
\begin{equation*}
    Q(\lambda) \equiv q(\lambda) \cdot \sigma, 
\end{equation*} to first order in $\varepsilon$  \cite{nielsengeometriclowerbound, Brandt2009}, we have
\begin{equation}
    \exp\big\{-i\big[Q(\lambda) + \varepsilon\dot{Q}(\lambda)\big]\big\} = \exp\big[-i \varepsilon H(\lambda)\big] \exp\big[-i Q(\lambda) \big]  \, , 
\end{equation} where the Hamiltonian $H(\lambda)$ is given by the series
\begin{equation}
    H(\lambda) = \sum_{k=0}^{\infty} \frac{ (-i)^{k} }{(k+1)!} \, \textrm{ad}^{k}_{Q(\lambda)} \, \dot{Q}(\lambda) ,
    \label{changeofbasis}
\end{equation}
%and we introduced the notation
with $\textrm{ad}_{A} B := [A, \, B] $, and $\textrm{ad}^{k}_{A}$ is the composition of this operator with itself $k$ times.

\subsubsection{Quantum complexity - Pauli geodesics} \label{sub_pauli_geo}
%Let us connect the notions of Finsler structure and Lie group

Let $F$ be a Finsler structure on $SU(2^n)$. We say $F$ is \emph{right-invariant} if
\begin{equation*}
F(\mathds 1, y) = F\big(r_{U}(\mathds 1), \, r_{\ast,U}(y)\big)\, ,
\end{equation*}
where $y \in T_{\mathds 1}SU(2^n)$.  Also, we say that $F$ is \emph{Pauli-symmetric} if it is right-invariant and has the property that, locally, $F$ does not depend on the sign of the coordinates $h^{k}$ (induced by a $U$-adapted chart), but only on the absolute values $|h^{k}|$ of these coordinates \cite{nielsengeometriclowerbound}. Note that the condition of right-invariance implies that the metric $F$ is completely determined by its values on the tangent space at the identity, that is, for any $U\in SU(2^n)$ and $v\in T_{U}SU(2^n)$, we have
\begin{equation}
 \left\| v \right\|_{F, U}   = \|r_{\ast,U^{-1}}(v)\|_{F, \mathds{1}}\, ,
\end{equation}
or, equivalently, in components
\begin{equation}\label{norm_F_U=norm_F_1}
 \left\|  \sum_{k=1}^{4^{n}-1} \, v^{k}   \, e^{\prime}_{k, U}\right\|_{F,U} = \left\| \sum_{k=1}^{4^{n}-1} \,  v^{k}   \,  e_{k,\mathds{1}} \right\|_{F, \mathds{1}} ,
\end{equation}
where $(v^{1}, \cdots ,v^{4^{n}-1}) \in \mathbb{R}^{4^{n}-1}$ are the components of $v \in T_USU(2^n)$ induced by $U$-adapted coordinates.

In this work, we are particularly interested in metrics as in references \cite{geometryofquco, nielsengeometriclowerbound, Brandt2009}, where, in terms of $U$-adapted coordinates, at $U$ they are given by
\begin{equation}\label{g|U}
g_{kl}|_U = \delta_{kl} \, w_{k},
\end{equation} where $w_{k} \in \mathbb{R}$ introduces \emph{weights} or \emph{penalties} depending on the number of terms different from the $2\times2$ identity that are present in the tensor product \eqref{sigma_k} of the generalized Pauli matrix $\sigma_{k}$. Akin to the first geometric formulation for quantum complexity \cite{quantumcompasgeo}, we adopt a metric that puts weights on operations involving 3 or more qubits in the Hamiltonian
\begin{equation}
w_{k} = \begin{cases}
1,\, \text{for operations on 1 or 2 qubits;}\\ 
4^{2n} , \,\, \text{for operations on 3 or more qubits.}
\end{cases}
\end{equation}
In our notation, we have
\begin{equation}
    \left\| \sum_{k=1}^{4^{n}-1} \,  h^{k}   \, e^{\prime}_{k,U}  \, \right\|_{F,U}^2 = \sum_{k=1}^{4^{n} - 1}   \,  w_{k}  \, \left| h^{k} \right|^{2}.
    \label{finslermany}
\end{equation} 
We note that the metric in equation \eqref{finslermany} is right-invariant, since its dependence on the components $h^{k}$ induced by the $U$-adapted charts is the same at any point $U \in SU(2^{n})$ \cite{nielsengeometriclowerbound}. Additionally, we explicitly see that this metric is Pauli-symmetric since it depends only on the modulus of the coefficients $h^{k}$ \cite{nielsengeometriclowerbound}.

We emphasize that %this definition
the expression \eqref{g|U} of the metric is given by coordinates centered at $U$, 
%of the metric is given at a single point, typically on the tangent space at the identity,
and should be understood as a point-wise specification, however, due to right-invariance, specifying the metric at %the identity
any point
fully determines it across the entire Lie group via \eqref{norm_F_U=norm_F_1}. %That is, the metric is not being defined over an open set, but rather assigned individually at each point
But we are not specifying explicitly how the metric varies along the coordinate chart, %.
%. This does not mean %, however, that geometric notions such as 
for instance the curvature does not vanish. In fact, on an open set $\mathcal{U}$ around the identity, the components of the Hessian matrix $g_{lk}(x)$ become functions of the coordinates $x$ only, thus characterizing a Riemannian geometry and a presentation of how to compute such curvature-related concepts in this context can be found in \cite{Brandt2009}.

Again, it is worth to mention that fixing the coordinate system is convenient for the geodesic computation. The conversion between the functional form of this $F$ into Pauli coordinates is generally not trivial. This change of basis is governed by equation \eqref{changeofbasis}, so that the coordinates $h(\lambda)$ depend on both $q(\lambda)$ and $\dot{q} (\lambda)$ \cite{geometryofquco}. However, if at every point on this curve we ensure that $[Q(\lambda), \, \dot{Q}(\lambda)] = 0$, then $h(\lambda) = \dot{q}(\lambda)$, and the problem is radically simplified. We will explore this fact and solve the Euler-Lagrange equations for a class of Pauli-symmetric Finsler metrics, such as the standard metric for quantum complexity in equation \eqref{finslermany}. To do so, we will introduce the so-called stabilizer subgroup and use an isometry of these metrics to derive the geodesic equations. These are the so-called Pauli geodesics \cite{nielsengeometriclowerbound}.

%In the context of the Lie group $SU(2^n)$, an isometry of a metric is a smooth function that maps elements of the Lie group to other elements within the same group, and after applying this function, the length of all curves remains the same. We consider the following isometry for this type of metric:
In our context of Pauli-symmetric Finsler metrics on $SU(2^n)$, we consider the following isometries:
\begin{equation}
    \varphi_\sigma(V) = \sigma V\sigma^\dagger\, ,
\end{equation}
where $\sigma$ is any element of the \emph{Pauli group} $\mathcal P = \{\pm \sigma_k, \pm i\sigma_k: k=0,...,4^n-1\}$,
\begin{comment}
$$\begin{array}{cccc}
   \varphi_{k}: & \textrm{SU}(2^{N}) &\xrightarrow[]{} &  \textrm{SU}(2^{N}) \\[1ex]
     &  V &\mapsto &   \varphi_{k}(V) \equiv \Tilde{\sigma}_{k} \hspace{0.05cm} V \hspace{0.05cm} \Tilde{\sigma}^{\dagger}_{k}.
\end{array}$$
\end{comment} 
which is known as the adjoint action of the Pauli group \cite{nielsengeometriclowerbound}. %Here, $\Tilde{\sigma}_{k}$ denotes that the Pauli matrices are multiplied by  real $\pm1$ or imaginary $\pm i$ scalar factors, to obtain a group structure, also including the identity matrix $\sigma_{0}$.
For a review of this group, interested readers can take a look at Chapter 10 of Nielsen and Chuang's book \cite{nielsen_chuang}. 

Now, we are interested in studying a subgroup $S$ of the entire Pauli group, such that, when we consider the adjoint action of $S$, the endpoint of a prefixed curve is not altered. The group $S$ is the \emph{stabilizer group} of this point \cite{nielsen_chuang,nielsengeometriclowerbound}. The initial point of such \ad{a} curve will be taken as the identity, which also remains unchanged. %\sub{Furthermore, we know that the adjoint action is an isometry. Therefore, a geodesic joining $\mathds 1$ to some point $U_f$ remain fixed by the action of the stabilizer of $U_f$. \,}
{Moreover, since the adjoint action of the Pauli group is an isometry, any geodesic curve $U:[0,1] \to SU(2^{n}), $ connecting $U(0) = \mathds{1}$ to a final point $U(1) = U_f$, remains invariant under the action of the stabilizer of $U_f$.}

%Therefore, as the initial and final points remain unchanged and this function does not alter lengths, %the geodesic is not affected by this action.

%The elements of $S$ are the independent elements of the Pauli group that commute and form a subgroup of the Pauli matrices.

%\sub{This isometry property under the action of the Pauli group enables us to determine the geodesic curve that connects the identity and some final point.  Assuming the final point of a curve $U$ is given by}
{This geodesic property under an isometric action allows us to solve the geodesic equation for a class of problems of Pauli-symmetric Finsler metrics, by imposing the following restriction. Let $X_f$ be such that}
\begin{equation*}
    U_f = \exp(-iX_{f})\,,
\end{equation*}
%\sub{and that we can express $X_{f}$}
{and assume that $X_{f}$ can be expressed}
    %in terms of the elements of a set $\mathcal{C}$, that is,
as a linear combination of some commuting set of Pauli matrices $\mathcal C =\{\sigma_{k_1},...,\sigma_{k_m}\}$, that is,
\begin{equation*}
    X_{f} = \sum_{l=1}^m \,  \, x_{f}^l \, \sigma_{k_l}\,.
\end{equation*}
%We can then construct a subgroup $S$ that stabilizes the operator at the final point by commutating with it. 
Then, the stabilizer subgroup of $U_f$ is the subgroup $S$ generated by $\mathcal C$. 
%Through this isometry, i
It can be shown \cite{nielsengeometriclowerbound} that if $U (\lambda)$ is a geodesic curve joining $\mathds 1$ to $U_f$, and $\sigma_{k} \notin S$, then
\begin{equation*}
x^{k}\big(U(\lambda)\big) = 0.
\end{equation*}
%Therefore, the isometry given by the adjoint action of the stabilizer group does not change the geodesic, but its symmetries allow us to study and determine the geodesics themselves.

%\sub{Noticing that}
{Therefore, } any candidate curve for such geodesic must have Pauli coordinates $q^{k}(\lambda)$ that are non-zero only in directions where $\sigma_{k} \in S$, and for these curves, it holds that
\begin{equation*}
    \big[q(\lambda) \cdot \sigma,  \, \dot{q} (\lambda)  \cdot \sigma \big] =0\,, 
\end{equation*}
so that
\begin{equation*}
\exp\big\{-i \big[q(\lambda) +\epsilon  \dot{q}(\lambda) \big] \cdot \sigma\big\} = \exp[-i \epsilon \dot{q}(\lambda) \cdot \sigma]  \exp[-iq(\lambda) \cdot \sigma].
\end{equation*}
That is, for these curves, $h(\lambda) = \dot{q}(\lambda)$%\sub{. Therefore,}
{, and} the coordinate changes are trivial, so we have the same vector components for both reference frames, with and without lines
\begin{equation*}
\sum_{k=1}^{4^{n}-1} \,  \dot{q}^{k}(\lambda)    \,  e_{k,U(\lambda)} =  \sum_{k=1}^{4^{n}-1} \,  \dot{q}^{k}(\lambda)   \, e^{\prime}_{k,U(\lambda)}.     
\end{equation*} 
Thus, by right invariance, we have that if two vectors at different points have the same components in $U$-adapted coordinates, then the norm of these vectors is the same, %But since the terms in the exponentials commute, the exponential separates, making the coordinate system changes trivial. In other words, we can remove the line from the vectors in equation \eqref{finslermany}, and we obtain
\begin{equation}
\left\|\,\sum_{k=1}^{4^{n}-1} \,  \dot q^{k}(\lambda)  \,  e_{k,\mathds 1} 
\,  \right\|_{F,\mathds 1}  =  \left\|\,\sum_{k=1}^{4^{n}-1} \,  \dot q^{k}(\lambda)   \,  e_{k,U(\lambda)} \, \right\|_{F,U(\lambda)}.
    \label{Fpauli}
\end{equation} %Therefore, we defined the metric in the Hamiltonian representation and obtained a way to calculate the norm of vectors in the Pauli representation, in the $\mathrm{S}$ frame.
Thus, from equation \eqref{Fpauli}, $F \big(0, p\big) = F\big(q, p\big).$ Hence
\begin{equation}
\dfrac{\partial \, F^{2}\big(q, p\big)}{\partial q^{k}} = 0.
 \label{delF2delq}
\end{equation}

%The function $F$ written in coordinates maps an open set in $\mathbb{R}^{2 \dim(M)}$ into an open set in $\mathbb{R}$; therefore, 
By writing $F$ in coordinates, we can use tools from variational calculus to find the geodesics. %For Finsler Pauli-symmetric metrics, if the final Hamiltonian $X_{f}$ can be written in terms of matrices in the set $S$, the geodesic is simply, as stated in \cite{nielsengeometriclowerbound}: $$U_{g}(\lambda) = \exp(-i X_{f} \lambda).$$ These are known as the Pauli geodesics.
Considering the Euler-Lagrange equations for same geodesic $U(\lambda)$ as before, in the $k$-th direction we have
\begin{equation*}
    \dfrac{d}{d \lambda} \left[ \dfrac{\partial\, F^{2}\big(q(\lambda), \dot{q} (\lambda) \big)}{\partial \dot{q}^{k}} \right]  + \dfrac{\partial \,  F^{2}\big(q(\lambda), \dot{q}(\lambda) \big) }{\partial q^{k}} = 0,
\end{equation*}
where $\sigma_{k} \in S$. Here, as in \cite{nielsengeometriclowerbound}, we substitute the minimization of $F$ with $F^{2}$ since these solutions are equivalent for length calculations. It follows from equation \eqref{delF2delq} that
\begin{align*}
0 &=  \dfrac{d}{d \lambda} \left[ \dfrac{\partial \, F^{2} }{\partial \dot{q}^{k}}\right]  \\[1ex]
&=  \sum_{l=1}^{4^{n}-1}\, \dfrac{d q^{l}(\lambda)}{d \lambda}  \, \dfrac{\partial \, F^{2}}{\partial q^{l} \, \partial \dot{q}^{k}} +  \, \dfrac{d \dot{q}^{l}(\lambda)}{d \lambda}  \,  \dfrac{\partial \, F^{2}}{\partial \dot{q}^{l}   \partial \dot{q}^{k}}  \\[1ex]
&=  \sum_{l=1}^{4^{n}-1} \,  \dfrac{d \dot{q}^{l}(\lambda)}{d \lambda}  \, \dfrac{\partial \, F^{2} }{\partial \dot{q}^{l}   \partial \dot{q}^{k}}   \\[1ex]
&= \sum_{l=1}^{4^{n}-1}  \, \dfrac{d \dot{q}^{l}(\lambda)}{d \lambda}  \, g_{lk}\big(q(\lambda), \dot{q}(\lambda)\big),
\end{align*} by the definition of the Hessian matrix of equation \eqref{Hessian matrix}. Applying its inverse to both sides, it follows that
\begin{equation*}
    \dfrac{d \dot{q}^{k}(\lambda)}{d \lambda}  = 0,
\end{equation*}
therefore $\dot{q}^{k}(\lambda) = c^{k}$ and $q^{k}(\lambda) = c^{k} \lambda$, for certain constants $(c^{1}, \cdots , c^{4^n -1} ) \in \mathbb{R}^{4^n -1}$. It follows from the boundary conditions that the geodesic is the curve:
\begin{equation}
    U(\lambda) = \exp(-i X_{f} \lambda).
\end{equation}

As mentioned early, these are called Pauli geodesics \cite{nielsengeometriclowerbound} since they arise from symmetries and isometries of the metric under the adjoint action of the Pauli group, together with the condition that the target point $U_f$ is generated by a Hamiltonian $X_f$ composed of mutually commuting generalized Pauli operators. This problem is particularly instructive, as it highlights the main challenges present in the general case while also demonstrating how they can be overcome for this simple case. In this specific setting, the variational approach allows for an exact solution of the geodesic curve, providing valuable insights into the geometry of quantum complexity.

\subsection{Barren Plateau Theory}
Quantum processors in the Noisy Intermediate-Scale Quantum (NISQ) era currently lack the necessary scale and coherence to support fault-tolerant quantum algorithms. As a result, they are not yet capable of independently solving many classes of complex computational problems. In this context, Variational Quantum Algorithms (VQAs) have emerged as a promising class of quantum applications, even within the limitations of the NISQ era \cite{zimboras2025myths, cerezo2021variational}. These algorithms combine quantum and classical computing resources by employing a classical optimizer to train a parameterized quantum circuit, in a manner analogous to training neural networks.

The essence of VQA is an iterative optimization loop designed to determine the optimal parameters that govern a specified quantum procedure. Each iteration begins by preparing an initial state, typically $|0\rangle^{\otimes n}$, on an $n$-qubit Hilbert space $\mathcal{H} = (\mathbb{C}^2)^{\otimes n} $. This state is then evolved by a \emph{Parameterized Quantum Circuit} (PQC) known as an \emph{ansatz}, represented by a unitary operator $U(\boldsymbol{\theta})$ described through some real parameters condensed in $\boldsymbol{\theta}$. Finally, a measurement of the output state is used by a classical optimizer to propose improved parameters for the next iteration. The unitary ansatz $U(\boldsymbol{\theta}) $ is typically constructed from a sequence of layers: for a fixed set of Hermitian operators $\mathcal G = \{H_1,..., H_m\}\subset i\,\mathfrak{u}(2^n)$, we take $\boldsymbol\theta = (\boldsymbol \theta_1,...,\boldsymbol \theta_L)$, for $\boldsymbol \theta_l = (\theta_l^1,...,\theta_l^m)\in \mathbb R^m$, and set
\begin{equation}\label{u(theta)}
    U(\boldsymbol{\theta}) = \prod_{l=1}^LU_l(\boldsymbol\theta_l) , \ \ U_l(\boldsymbol\theta_l)=\prod_{k=1}^m \exp(i \, \theta_l^k \, H_k) \ad{\, .}
\end{equation}
In this structure, the overall ansatz is composed of $L$ sequential unitary layers. When the number of these layers, $L$, is large, the architecture is often referred to as a \emph{Deep Parameterized Quantum Circuit}. {In addition, we suppose there is a systematic method to construct a PQC for each $n\in \mathbb{N}$, in a way that it is possible to talk about the asymptotic behavior for a large number of qubits.} 

This construction enables the study of parameterized quantum circuits through the powerful mathematical tools of compact Lie groups and their corresponding Lie algebras.
In particular, the set of Hermitian generators defines the \emph{Dynamical Lie Algebra} (DLA), denoted by $\mathfrak{g}$, as the Lie algebra generated by $i\mathcal G$:
\begin{equation}
\label{DLA}
    \mathfrak{g} = \langle i\mathcal{G} \rangle_{\mathrm{Lie}} \text{.}
\end{equation}
Since $\mathfrak{g} \subseteq \mathfrak{u}(2^n)$, it forms a {reductive Lie algebra, so it can be decomposed as in \eqref{decomp_red_lie_alg}. For the remaining of this section, the compact Lie group given by exponentiation of $\mathfrak g$ will be denoted by $G$, and we refer to the set of all unitary operators $U(\boldsymbol \theta)$ of the form \eqref{u(theta)} as an \emph{ensemble} $\mathcal E_L$ in $G$.

{To justify the name ``ensemble'', we have to consider a probability distribution on $\mathcal E_L$. If, by any chance, $\mathcal E_L$ is itself a Lie group, its Haar measure is a very natural probability to consider -- it happens, for example, when $\mathcal G$ is comprised by mutually commuting operators, so that $\mathfrak g$ is abelian and $\mathcal E_L = G$ is a torus. However, in general, $\mathcal E_L$ can be far from an actual group, and a natural notion of ``uniform probability'' seems to be lacking. Indeed, the probability distribution to be considered is induced by the parametrization (cf. \cite{larocca2022diagnosing}), reflecting the factual process of constructing the circuit. First, note that, for any $j\in \{1,...,m\}$, the map $\theta\mapsto e^{i\theta H_j}$ on $\mathbb R$ has a fundamental period, say, $\tau_j>0$. Thus, each layer can be parameterized by the $m$-torus
\begin{equation}
    T^m = (\mathbb R \mod \tau_1)\times...\times (\mathbb R \mod \tau_m)\, ,
\end{equation}
where $\alpha,\beta\in \mathbb R$ represents the same point in $(\mathbb R\mod \tau_j)$ if $\alpha-\beta = a\,\tau_j$ for some $a\in\mathbb Z$. Hence we restrict the parametrization \eqref{u(theta)} of $\mathcal E_L$ to the $(mL)$-torus $(T^m)^L$. Then we take the probability measure $\mu_L$ on $\mathcal E_L$ induced from the Haar measure of $(T^m)^L$. In precise words, $\mu_L$ is the \emph{pushforward of the Haar measure} by the surjection $(T^m)^L\to\mathcal E_L$. That is, if $f:\mathcal E_L\to \mathbb C$ is any integrable function, then}
{\begin{equation}
\begin{aligned}
    & \int_{\mathcal E_L}d\mu_L(U)f(U) = \int_{(\mathbb T^m)^L}d\boldsymbol\theta f(U(\boldsymbol\theta))\\
    & =\int_{T^m}d\boldsymbol\theta_L...\int_{T^m}d\boldsymbol\theta_1f(U(\boldsymbol\theta_1,...,\boldsymbol\theta_L))\, ,
    \end{aligned}
\end{equation}}
{where}
\begin{equation}
    {\int_{T^m}d\boldsymbol\theta_l = \dfrac{1}{\tau_1...\tau_m}\int_0^{\tau_1}d\theta_l^1...\int_0^{\tau_m}d\theta_l^m\, .}
\end{equation}
{Again, when $\mathcal G$ is comprised by mutually commuting operators, $\mu_L$ coincides with the Haar measure on $\mathcal E_L = G = T^m$.}

{Then we can quantify how much $\mu_L$ is distant from the Haar sample on $G$ by considering, for each $k\in \mathbb N$, the operators $\mathcal M_{\mathcal E_L}^{(k)}:\mathcal B(\mathcal H^{\otimes k})\to \mathcal B(\mathcal H^{\otimes k})$, }
\begin{equation}\label{k-stat-moment-op}
   {\mathcal M_{\mathcal E_L}^{(k)}(M) = \int_{\mathcal E_L}d\mu_L(U) U^{\otimes k} M (U^\dagger)^{\otimes k}\, ,}
\end{equation} 
{called the \emph{$k$-th statistical moment operator}. When the circuit is sufficiently deep, it can approximate a \emph{$t$-design}, meaning that the first $t$ statistical moment operators approximates the respective statistical moment operators of $G$ defined in the same vein of \eqref{k-stat-moment-op}, just substituting $\mathcal E_L$ and $\mu_L$ by $G$ and its Haar measure (cf. e.g. \cite{dankert, ragone2024lie}).}

\begin{comment}
\ex{Note that, for any $j\in \{1,...,m\}$, the map $\theta\mapsto e^{i\theta H_j}$ on $\mathbb R$ has a fundamental period, say $\tau_j>0$. Thus, each layer can be parameterized by the $m$-torus}
\begin{equation}
    \ex{T^m = (\mathbb R \mod \tau_1)\times...\times (\mathbb R \mod \tau_m)\, ,}
\end{equation}
\ex{where $\alpha,\beta\in \mathbb R$ represents the same point in $(\mathbb R\mod \tau_j)$ if $\alpha-\beta = a\,\tau_j$ for some $a\in\mathbb Z$. Hence the parametrization \eqref{u(theta)} of $\mathcal E_L$ can -- and will -- be restricted to the $(mL)$-torus $(T^m)^L$.}
\end{comment}

This formulation provides the necessary tools to rigorously analyze the trainability of structured parameterized quantum circuits. A key aspect of this analysis involves studying the landscape of the loss (or cost) function, which for many VQAs can be framed in the following form \cite{cerezo2021cost,larocca2022diagnosing}:
\begin{equation}
\ell_{\boldsymbol{\theta}}(\rho, O) = \mathrm{Tr} \left[ U(\boldsymbol{\theta}) \rho U^{\dagger}(\boldsymbol{\theta}) O \right]{\, ,}
\end{equation}
where $O$ is a Hermitian operator and $\rho$ is the initial state. The geometric properties of this loss landscape are critical to the performance of a VQA.

Among the principal challenges in this landscape is the \emph{barren plateau} (BP) phenomenon, where the loss function and its gradients become exponentially concentrated around their mean value as the number of qubits, $n$, increases. Significant research has been dedicated to understanding the sources of barren plateaus, with several factors identified as primary causes. These include the expressiveness of the ansatz \cite{holmes2022connecting}, high entanglement in the initial state \cite{ortiz2021entanglement}, the non-locality of the measurement operator, and the presence of hardware noise \cite{ragone2024lie,cerezo2023does}.

A key quantitative indicator of a barren plateau is the variance of the loss function. If the variance vanishes exponentially with the number of qubits, an exponential number of measurements is required to resolve a gradient, rendering the algorithm untrainable \cite{cerezo2021variational}. Conversely, a polynomially scaling variance indicates the system is trainable. This challenge is rooted in the practical estimation of expectation values, which relies on a finite number of measurement shots and is subject to statistical uncertainty. Detecting an exponentially small signal (i.e., the gradient) amidst this shot noise requires an exponentially large number of samples, making the VQA inefficient and non-scalable.

The variance of the loss function is calculated in \cite{ragone2024lie} for some special cases: under the assumption that $\mathcal E_L$ is a $2$-design, if $\rho$ or $O$ lies in $\mathfrak g$, the variance of $\ell_{\boldsymbol\theta}(\rho, O)$ is given by
\begin{equation}
    \label{eq:BP_general}
    \text{Var}_{\boldsymbol\theta}[\ell_{\boldsymbol\theta}(\rho, O)] = \sum_{k=1}^{d} \frac{\mathcal{P}_{\mathfrak{g}_k}(\rho)\mathcal{P}_{\mathfrak{g}_k}(O)}{\operatorname{dim}(\mathfrak{g}_k)}{\, ,}
\end{equation}
where each $\mathfrak g_k$ is a simple ideal of the decomposition \eqref{decomp_red_lie_alg} of $\mathfrak g$, and $\mathcal P_{\mathfrak g_k}(H)$ is called the \emph{$\mathfrak g_k$-purity} of the Hermitian operator $H$, given by
\begin{equation}
    \mathcal P_{\mathfrak g_k}(H) = \tr(H_{\mathfrak g_k}^2)\, ,
\end{equation}
for $H_{\mathfrak g_k}$ being the projection of $H$ onto $\mathfrak g_k$. For pedagogical reasons, let's consider only the case where the DLA is a simple Lie algebra, so the sum in \eqref{eq:BP_general} reduces to a single term:
\begin{equation}
    \label{eq:BP_simple}
    \text{Var}_{\boldsymbol\theta}[\ell_{\boldsymbol\theta}(\rho, O)] = \frac{\mathcal{P}_{\mathfrak{g}}(\rho)\mathcal{P}_{\mathfrak{g}}(O)}{\operatorname{dim}(\mathfrak{g})}.
\end{equation}

The dimension of the Dynamical Lie Algebra (DLA), denoted as $\operatorname{dim}(\mathfrak{g})$, provides a direct algebraic measure of an ansatz’s expressibility. An ansatz is considered expressible if the associated quantum circuit is capable of uniformly exploring the entire space of quantum states. When $\operatorname{dim}(\mathfrak{g})$ is small, the group $G$ generated by the circuit in the deep limit forms a low-dimensional submanifold of the unitary group $U(2^n)$. As a result, the accessible state space is significantly restricted. Conversely, when $\operatorname{dim}(\mathfrak{g})$ is large (e.g., exponential in $n$), the group $G$ can approximate the full unitary group $U(2^n)$ more closely. This observation leads to an important implication for trainability: deep parameterized quantum circuits whose DLA dimension scales as $\Omega(b^n)$ for some constant $b \geq 2$ are subject to barren plateaus even for good choices of observable or initial quantum state. %\footnote{\ex{This follows directly as a corollary of result \eqref{eq:BP_general}.}} \ad{as can be inferred from {\eqref{eq:BP_simple}}.}

{The $\mathfrak g$-purity of a state $\rho$ gives a measure of entanglement} not with respect to a spatial partitioning of qubits, but relative to the subspace of observables defined by the DLA: a small value of $\mathcal{P}_{\mathfrak{g}}(\rho)$ means a high degree of generalized entanglement, indicating that the state has significant components outside the subspace defined by $\mathfrak{g}$. This is a critical point: a state that is highly entangled with respect to the DLA can induce a barren plateau. This can occur even if the circuit architecture itself generates a DLA with $\operatorname{dim}(\mathfrak{g}) \in \mathcal{O}(\text{poly}(n)$. This highlights that the trainability of a variational quantum circuit is influenced not only by the expressibility of the circuit itself but also by intrinsic properties of the quantum state being prepared.

Similarly, for the measurement operator $O$, the $\mathfrak{g}$-purity provides a notion of \emph{generalized locality}: in this framework, an operator's ``locality'' is determined by its alignment with the DLA of the circuit. An operator is considered \emph{generalized-local} if it is well-aligned with or lies within the DLA. In the ideal case where $O \in i\mathfrak{g}$, its projection is itself , $O_{\mathfrak{g}} = O$, which maximizes the value of $\mathcal{P}_{\mathfrak{g}}(O)$ and, consequently, its contribution to the variance. Conversely, an operator is \emph{generalized-nonlocal} if it is poorly aligned with the DLA, resulting in a small projection $O_{\mathfrak{g}}$ and a correspondingly small $\mathfrak{g}$-purity. Just like the state's purity, if an operator is highly generalized-nonlocal, its $\mathfrak{g}$-purity can become exponentially small, creating a barren plateau independently of the circuit's expressiveness or the initial state's properties. This reframes the conventional wisdom that only ``global'' observables cause barren plateaus, showing that the crucial factor is the operator's relationship with the circuit's underlying algebraic structure.

\section{Discussion}

This work %demonstrated
explored 
the profound utility of Lie group and Lie algebra theory as an unifying framework for analyzing complex problems in quantum computing. 
%By moving beyond a simple matrix description of quantum gates and 
By reviewing some recent works on theoretical quantum computing,
embracing the geometric and algebraic structure of the underlying $SU(2^n)$ Lie group, we have provided %a rigorous foundation 
a pedagogical introduction to %address challenges
understand fundamental tools required for studying
%in both variational quantum algorithms trainability and computational complexity.
the geometry of quantum complexity and variational quantum algorithms trainability.

In the context of quantum complexity, we adopted a geometric perspective, defining the cost of a quantum operation as the length of a geodesic path on the $SU(2^n)$ manifold. 
By employing a right-invariant Finsler metric designed to penalize physically difficult multi-qubit interactions, we have described a concrete method for quantifying circuit complexity. 
A significant result of this approach was the identification of ``Pauli geodesics''. 
We showed that for a target unitary composed of commuting Pauli operators, the optimal synthesis path is a simple straight line in the Lie algebra. This result exemplifies the calculation of complexity for a simple class of quantum operations.
%This result radically simplifies the calculation of complexity for this important class of quantum operations.

In our analysis of VQAs, we pointed out a direct link between the algebraic properties of a circuit's ansatz and the practical issue of trainability. 
%We showed that the DLA, generated by the circuit's Hamiltonians, is a key determinant of performance. 
The emergence of barren plateaus was explicitly tied to the dimension of the DLA, where an %DLA with a dimension that grows exponentially
exponential growth
in %the number of qubits
the dimension of this algebra leads to an exponentially vanishing gradient variance. 
Furthermore, we highlighted the concept of $\mathfrak g$-purity, demonstrating that barren plateaus can also be induced by the choice of initial state or measurement operator, specifically when they exhibit high ``generalized entanglement'' or ``generalized non-locality'' with respect to the DLA. 
This provides a clear algebraic explanation for the challenges in training deep parameterized quantum circuits.

Taken together, these two applications demonstrate that the abstract framework of Lie theory offers concrete and quantitative tools for quantum computer scientists. Concepts such as geodesics on the $SU(2^n)$ manifold, the dimension of the DLA, and the purity of a quantum state relative to that algebra are far from being mere mathematical curiosities; they are closely connected to practical performance indicators, such as algorithm trainability and operational efficiency. Ultimately, the integration of differential geometry and Lie theory into the study of quantum computation are essential for addressing key challenges in the field. This line of research opens fertile ground for the development of more powerful and robust quantum technologies.

\begin{acknowledgments}
%P.A. and L.M. acknowledge the Simons Foundation (Grant Number 1023171, RC). L.M. also thanks Rafael Chaves and Tailan Sarubi for their encouragement and insightful discussions that motivated this work. G.A. acknowledges financial support from the Coordenação de Aperfeiçoamento de Pessoal de Nível Superior (CAPES), project number 88887.901613/2023-00.
P.A. acknowledges Fundação Norte-Rio-Grandense de Pesquisa e Cultura (Funpec) for financial support.  G.A. acknowledges Reginaldo de Jesus Napolitano for valuable discussions, conceptual clarifications and help with some calculations. L.M. acknowledges the Simons Foundation (Grant Number 1023171, RC) for financial support, and Tailan Sarubi and Rafael Chaves for encouraging the writing of this paper. This study was financed in part by the Coordenação de Aperfeiçoamento de Pessoal de Nível Superior – Brasil (CAPES) – Finance Code 001. %G.A. acknowledges financial support from the Coordenação de Aperfeiçoamento de Pessoal de Nível Superior (CAPES), project number 88887.901613/2023-00. 
\end{acknowledgments}

\bibliography{refs.bib}

\end{document}